# A Scalable Microarchitecture for Efficient Instruction-Driven Signal Synthesis and Coherent Qubit Control

Nader Khammassi, Randy W. Morris, Shavindra Premaratne, Florian Luthi, Felix Borjans,
Satoshi Suzuki, Robert Flory, Linda Patricia Osuna Ibarra, Lester Lampert, Anne Y. Matsuura

Intel Labs, Intel Corporation, Oregon, USA
Email: nader.khammassi@intel.com

**Abstract**

Execution of quantum algorithms requires a quantum computer architecture with a dedicated quantum instruction set that is capable of supporting translation of workloads into actual quantum operations acting on the qubits. State-of-the-art qubit control setups typically utilize general purpose test instruments such as arbitrary waveform generators (AWGs) to generate a limited set of waveforms or pulses. These waveforms are precomputed and stored prior to execution, and then used to produce control pulses during execution. Besides their prohibitive cost and limited scalability, such instruments suffer from poor programmability due to the absence of an instruction set architecture (ISA). Limited memory for pulse storage ultimately determines the total number of supported quantum operations. In this work, we present a scalable qubit control system that enables efficient qubit control using a flexible ISA to drive a direct digital synthesis (DDS) pipeline producing nanosecond-accurate qubit control signals dynamically. The designed qubit controller provides a higher density of control channels, a scalable design, better programmability, and lower cost compared to state-of-the-art systems. In this work, we discuss the new qubit controller's capabilities, its architecture and instruction set, and present experimental results for coherent qubit control.

## 1   Introduction

Quantum computing is an emerging technology with the potential to address many intractable classical problems more efficiently [2][3]. Quantum algorithms are expressed as a set of quantum gates or operations that involves qubit initialization, manipulation, and measurement of qubits. One or more such measurement outcomes are then processed to compute the result of the workload. While the approach can be generalized to different qubit types, in this work, we show how we can use our qubit controller to control single-electron spin qubits in gate-defined semiconductor quantum dots (QDs) [9][12].

In the case of spin qubits, quantum operations correspond to a set of RF and/or DC pulses applied to the qubits to control their states. In addition to qubit pulsing, qubit measurement involves digitizing and processing of the signal received back from the qubit to determine its state. Qubit control setups are often composed of various signal generators to produce the control pulses, and an acquisition system such as lock-in amplifiers to process and digitize the readout signals. An alternative approach that has recently been explored is to utilize cryogenic CMOS control chips and aims to generate signals for qubit control while providing high integration at low temperature [16][17] such as the Horse Ridge cryo-controller chip [15].

In the next few paragraphs, we identify the limitations observed in state-of-the-art qubit control systems and we list the contributions of this work in addressing those limitations.

### 1.1   Limitations of State-of-the-art Qubit Control Systems

Figure 1 shows how a typical state-of-the-art qubit control setup [4][5][11] might utilize an AWG to produce pulses. The pulses are precomputed and stored in the AWG memory prior to the execution. A codeword is assigned to each pulse in the memory and is used by the AWG to fetch and produce the analog pulse corresponding to the input via a simple lookup table (LUT).

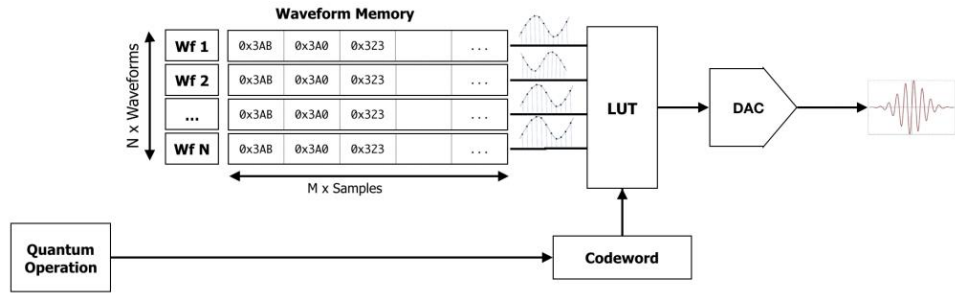

Figure 1. Standard AWG codeword-based approach relies on large memory to store pulses. The LUT will be used to map the waveform to the corresponding codeword.

Although the use of traditional AWG-based architectures enables convenient assembly of setups for controlling few-qubit systems they severely limit the performance when controlling many-qubit systems. Some of the major limitations include:

- Inefficient memory usage
- Waveform construction and uploading latency
- Absence of ISA limits the programmability and prevents efficient integration with a quantum compiler.

Since AWGs typically store the full waveform corresponding to each quantum operation, and there is a limited amount of memory, it is inevitable to have tradeoffs between competing characteristics (e.g. types of supported quantum operations vs. fineness/resolution of arbitrary qubit rotations vs. sampling rate of waveforms). Furthermore, especially in the case of spin qubits [9], short-duration fine-control single qubit operations will need to coexist with long-duration operations such as qubit readout; supporting both scenarios typically requires inefficient usage of AWG memory due to the need of various sampling rates.

The communication overhead is a significant bottleneck during qubit calibration and tune-up procedures since these routines require sweeping of various waveform parameters. As the sweeps require the construction and upload of numerous waveforms, the overall tune-up efficiency is also reduced when compensating for drifting system parameters. Moreover, this bottleneck can be even more significant when more frequent calibration is required.

There have been some studies on improving AWG programmability by introducing an external control interface with dedicated instruction sets [4][5]. However, the later approaches suffer from similar limitations to traditional instruments due to their reliance on the same memory-based AWG architecture. This ultimately limits the performance and capabilities of a qubit control system despite those improvements. *Figure 2* shows a typical spin qubit control setup using standard test instruments such as AWGs used for modulating microwave signals, Lock-In Amplifiers for readout and precision DC bias channels.

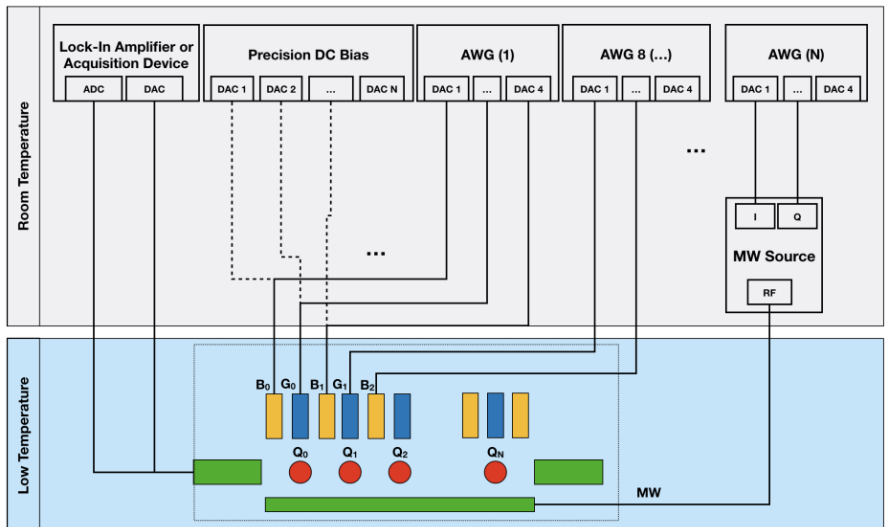

Figure 2. Conventional Qubit Control Setup for a Spin Qubit Chip.

## 1.2 Contributions of this work:

To address the limitations of the state-of-the-art AWG-based quantum control systems, we designed and implemented a new qubit controller. Our controller replaces all AWGs and Lock-In Amplifiers while providing instruction-based qubit manipulation and readout capabilities. The new system uses a novel instruction set and a dedicated qubit control architecture [13] specifically designed to control qubits efficiently while offering better scalability to control larger qubit systems.

As shown in *Figure 3*, the new instruction set enables dynamic synthesis of the pulses by driving a DDS pipeline. The use of DDS enables fast synthesis of the pulses on-the-fly obviating the need for a large memory to store sampled pulses. For a given quantum operation, the parameters of the pulses such as the frequency, phase, amplitude, shape, and duration are encoded in the instructions sent to the qubit controller. Once an instruction is fetched, decoded, and executed, a DDS pipeline uses those parameters to synthetize the pulse. Several digital signal processing (DSP) stages are responsible of shaping and filtering the pulses before sending them to the digital-to-analog converter (DAC). This allows the controller to achieve both high memory efficiency as well as enhanced programmability compared to state-of-the-art systems. Instructions are executed in many channels simultaneously to maintain precise synchronization and coherent control.

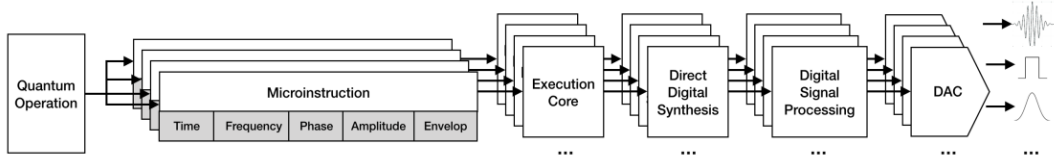

Figure 3. The flow for dynamic synthesis of waveforms using pulse parameters encoded in instruction bitfields.

Besides efficient control signal synthesis and high programmability, our microarchitecture integrates a qubit readout unit implementing different readout techniques namely charge sensing [18][19] and reflectometry-based readout [10][20]. The implementation allows performing fast statistics and correlation of the readout outcomes automatically in hardware, while enabling also digitizing raw signals for offline processing in software whenever needed.

Our qubit controller was implemented on a Stratix FPGA [21] driving 22 DAC Channels for radio-frequency (RF) and arbitrary waveforms, and 2 analog-to-digital converter (ADC) Channels for qubit readout. The system is housed in a 3U enclosure and managed by the host PC via a high-speed PCIe Link operating at 20 Gbps (see *Figure 4*). This link provides access to the control registers and the memory infrastructure (including DDR3 dynamic random-access memory (DRAM) and on-chip memories) through a dedicated direct memory access (DMA) controller. The resulting system is compact, offers a high channel density and can be integrated in standard racks used for qubit control setups. Moreover, our microarchitecture is designed to achieve high scalability through enabling a multi-controller mode where several qubit controllers can be used to control larger qubit systems while maintaining synchronization across the different controllers.

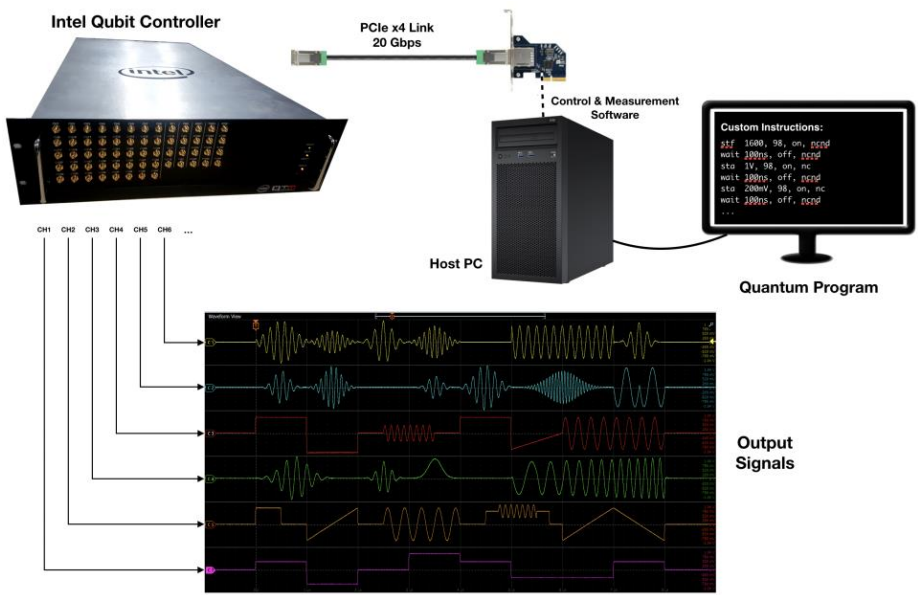

Figure 4. The qubit controller is managed by the host PC through a PCIe Link. Upon execution of instructions loaded onto the controller, the desired pulses are generated on all the channels with nanosecond-accurate timing.

In the next section we give an overview of the qubit controller microarchitecture. Section 3 is dedicated to presenting the instruction set. In Section 4, we describe the signal synthesis pipeline and its different features. Section 5 outlines the scalability of our design and how multiple controllers can be used to control larger qubit systems. Section 6 describes the qubit measurement unit and its capabilities. Finally, Section 7 will describe the integration of the qubit controller with the qubit chip, as well as demonstrating the experimental results.

## 2   System Architecture Overview

*Figure 5* schematically represents the system architecture of our qubit controller. The system consists of a modular hardware architecture that can be split into a few functional blocks:

- A PCIe Interface to provide a high-speed communication link with the control PC (20 Gbps).
- A DMA Controller and an interconnection fabric to provide access to various memories and control registers.
- A DDR3 DRAM Memory Controller with a large bandwidth to feed the instruction FIFOs and store readout signals.
- 22 parallel instruction execution cores driving parallel signal processing pipelines for generating, filtering, correcting the signals, and driving each DAC channel.
- A qubit measurement unit (MU) (in orange) implementing dedicated signal processing logic for qubit measurements, raw data acquisition, and statistics gathering.
- A clock distribution network with fine skew control to align the signals produced by different channels.
- A dedicated qubit controller driver to control the system through a PCIe driver.
- An application programming interface (API) implementing the different routines for communicating and controlling the qubit controller.
- A set of test tools designed for the test and validation of the various controller functionalities.

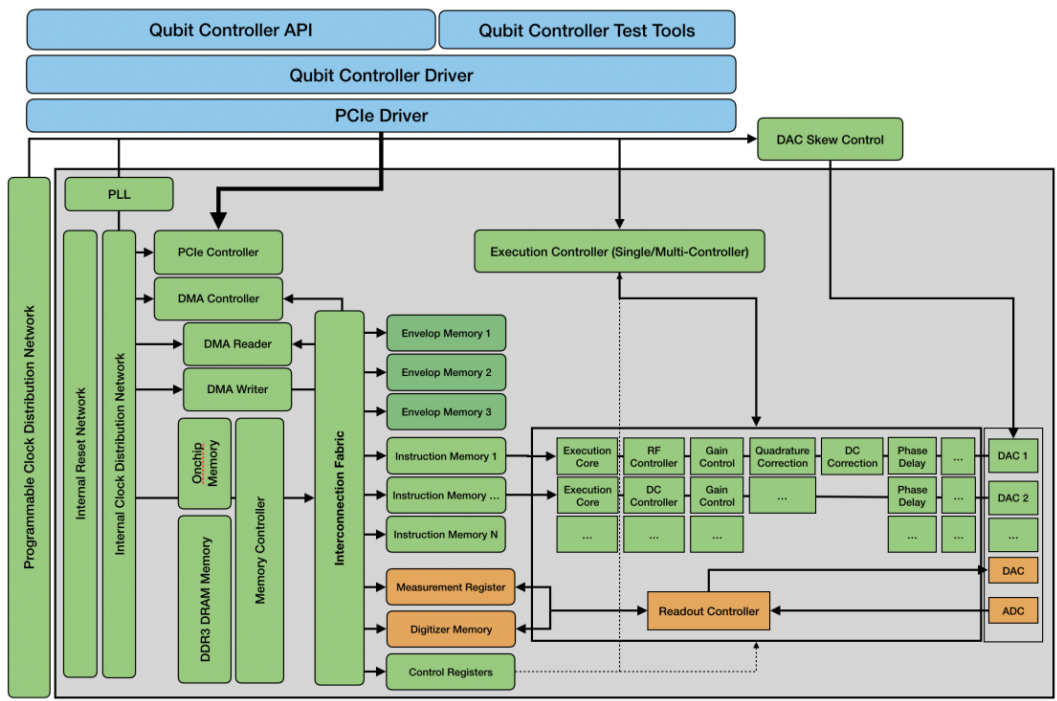

Figure 5. Qubit Controller Architecture Overview.

The API is used to connect to the qubit controller, configure it, load instructions into its memory, initiate the execution, and read back the results from the qubit measurement registers and the digitizer memory. The instructions are written to the DRAM through the DMA controller. When execution starts, the instructions are fetched by the various execution units, decoded, and executed. During execution, the decoded pulse parameters are used by the signal synthesis pipeline to produce the corresponding waveform and feed the samples to the DAC to produce the analog pulses (see *Figure 3*).

All the channels operate synchronously on the same clock to guarantee deterministic pulse timing and coherent qubit control. During qubit measurement, the readout instruction triggers the MU that sends pulses to the qubit and digitizes the readout signal received back from the qubits. The digitized signal is processed to determine the qubit state. The raw digitized signal is also automatically recorded and stored into the DRAM for offline processing and analysis by specialized users.

# 3 Instruction Set

Our qubit controller implements a new set of instructions [13] especially designed for efficient qubit control. The instruction set enables generation of pulses with precisely defined parameters such as frequency, phase, amplitude, shape, and duration. These instructions also enable precise nanoseconds-scale timing control and the control of the qubit measurement unit.

Instruction execution is fully pipelined and enables deterministic and cycle-accurate execution without dead time between consecutive pulses. In the current implementation on the Stratix FPGA, the clock cycle is set to 5ns and is adequate for the state-of-the-art qubit pulsing time scale. The design also supports higher clock speeds, if so desired. The shortest possible pulse duration and idling time is 5 ns, while the longest possible duration is on the order of many hours and thus nearly continuous, all while maintaining the same time resolution. The instructions can be classified functionally into three groups: *pulse synthesis*, *time control and synchronization* and *qubit readout instruction*.

## 3.1 Pulse Synthesis Instructions:

To precisely define pulse parameters, four unique instructions are utilized:

- **STA** (SeT Amplitude) Sets the amplitude of a target pulse on a given channel.
- **STF** (SeT Frequency) Sets the frequency of the pulse. The frequency is controlled via a Numerical Controlled Oscillator (NCO).
- **STP** (SeT Phase) Dedicated for controlling the phase shift relatively to the reference NCO phase.
- **STAP** (SeT Amplitude & Phase) Combines the functions of STA and STP to enable changes to both the phase and amplitude simultaneously. This instruction is specifically designed to accommodate a particular group of arbitrary single-qubit operations that are enacted via pulses with a unique phase and amplitude combination.

All of the above instructions include bitfields for specifying the duration of the pulse, the shape of the pulse as defined by an envelope identifier, the conditionality of the instruction on a specified qubit measurement outcome, and a flag for activating the target channel or deactivating the channel while changing the relevant parameters silently.

## 3.2 Time Control and Synchronization Instruction

**WAIT**: The wait instruction has two possible methods of operation.
- If the channel is turned off, this instruction is used to set idling times for a given number of cycles between pulses and to schedule instructions relative to other channels.
- If the channel is turned on, this instruction can be used to activate a channel using previously set pulse parameters (e.g. STA, STF, STP and STAP). This usage is particularly useful when generating very long pulses.

**SYNC**: Used to synchronize the execution with external instruments used in the qubit setup. It is designed for compatibility with existing qubit control setups that contain other instruments. The SYNC instruction blocks execution and lets the controller wait for a trigger from an external instrument. Upon receiving the trigger, execution resumes on the rising edge.

## 3.3 Readout Instruction

**RDO**: The readout instruction triggers the MU responsible for measuring the qubit states. The instruction specifies the duration of the measurement window and the qubit state discrimination parameters. Readout durations from nanoseconds to seconds are possible in the current controller implementation. When triggered, the readout unit generates quadrature pulses, processes the reflected pulses acquired by the ADC using a dedicated DSP pipeline to discriminate the qubit state and finally stores the outcome in the measurement register.

## 3.4 Conditional Execution

Instructions can be conditional and executed only if the specified bit of a measurement register is high or low. This can be used to implement several qubit operations involving conditional execution such as active qubit reset where the qubit is rotated actively to initialize it in a specific state, or error correction code implementation where qubit states are corrected based on the measurement outcome of ancilla qubits e.g. error syndromes.

Figure 6 shows the execution of a sequence of qubit controller instructions on an RF channel and the corresponding quadrature pulses at the output.

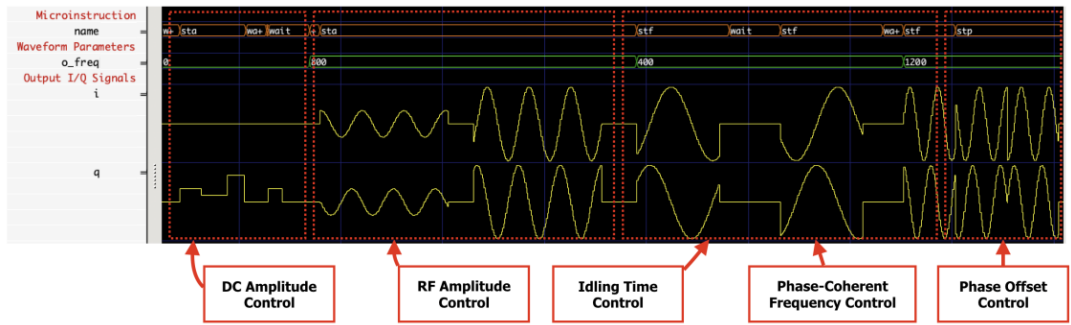

Figure 6. Microinstruction execution and the produced RF and DC waveforms.

## 4 Signal Synthesis

The qubit controller uses instruction-driven DDS to generate the desired signals and integrates several DSP stages designed to address several qubit control requirements. The qubit controller can generate three types of signals: RF, DC or arbitrary signal shapes.

On the RF signal path, an NCO locked to a low phase noise reference is responsible for generating a quadrature signal at the desired frequency. The phase shift, frequency, and amplitude are all controlled via instructions. Subsequent stages allow the application of arbitrary envelops to the RF signal, quadrature modulation correction, dc offset correction, predistortion, filtering, skew adjustment, etc.

On the DC/Arbitrary signal path, arbitrary voltage levels can be produced using the STA instruction, and this capability can be utilized to generate square pulses, isosceles trapezoidal pulses (by increasing the rise/fall times), or arbitrary waveforms (using direct sampling). Skew adjustment is also integrated in the processing pipeline.

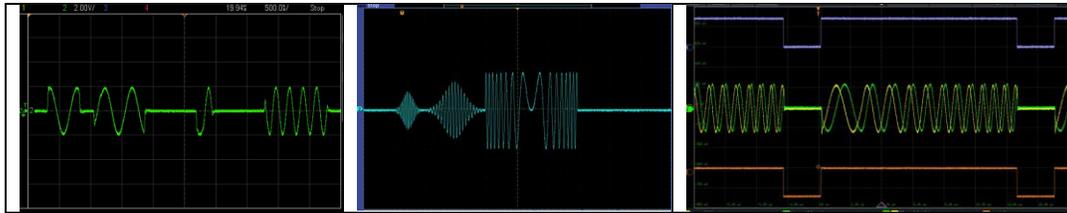

Figue 7. Example of actual synthetized waveforms: a) short pulsing with various frequencies, b) pulse shaping and c) adiabatic pulsing involving frequency sweeping and synchronized square pulses.

### 4.1 Quadrature Modulation Correction

RF driving for qubit control typically employs Single Side Band (SSB) modulation to modulate the microwave signal and produce a clean frequency spectrum devoid of undesired tones that can drive the qubits off-resonance [9] resulting in lower quantum operation fidelities. SSB modulation requires efficient quadrature modulation corrections (QMC) to focus the microwave power of the desired frequencies and eliminate undesired spurious signals. QMC is both a critical and tedious calibration routine when performed manually. Therefore the controller includes a quadrature modulation stage that can be used to correct the modulation and compensate for imperfections in external SSB modulators on-the-fly. Figures 8 and 9 show the frequency spectrum of a SSB modulated 15.025 GHz respectively before and after enabling the QMC.

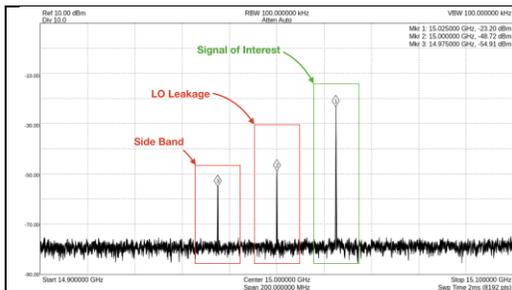 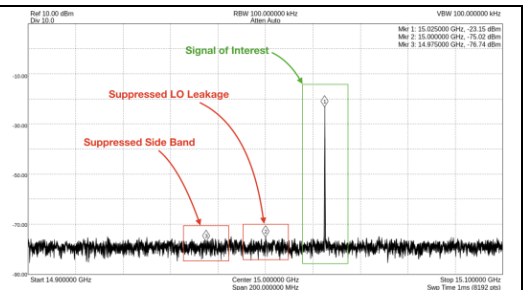

Figure 8 . Due to imperfections of Quadrature Modulators, absence of QMC results in several unwanted signals due to LO Leakage and unwanted sideband.

Figure 9 . Presence of QMC effectively suppresses the LO Leakage and the spurious side band by adjusting the DC offset and compensating for gain and phase unbalance.

### 4.2 Skew Correction

Qubit control setups involve many heterogenous components that include RF components, interconnection infrastructure, and various instruments. This results in various latencies in different signal paths and skew between channels. Our qubit controller integrates a digitally-controlled delay stage that enables the user to configure different delays and align the various control paths in the system.

### 4.3 Predistortion and Filtering.

Qubits are sensitive to imperfections in the control signals, therefore distortions that can occur through RF signal paths can significantly deteriorate the fidelity of the quantum operations [6][22]. To counteract this effect, our qubit controller integrates a predistortion block with a configurable digital filter on each RF channel and allows the user to compensate for arbitrary frequency responses caused by imperfect RF components on the signal path. This feature can be also used to perform arbitrary digital filtering.

## 5   Scalability: Multi-Controller System

As the number of qubits increases, the number of control channels increases dramatically, thus the scalability of our qubit control system is critical for supporting large qubit systems. Our controller has been designed to enable a multi-controller mode where several controllers work together to control larger qubit systems while maintaining synchronization between the different channels across different control units.

In the multi-controller mode, one unit is configured in the *Conductor* mode while the others are configured in the *Performer* mode. The *Conductor* is responsible for ensuring inter-unit synchronization as well as executing its own part of the workload. Instructions are loaded to each of the controller units, and control commands issued to all the *Performer* units. A dedicated clock distribution unit is responsible for adjusting and maintaining low skew between the different controller units. *Figure 10* gives an overview of a multi-controller system with one Conductor and N performers. The depicted multi-controller system is driven from one single control PC, but the architecture supports larger systems composed of a cluster of control PCs.

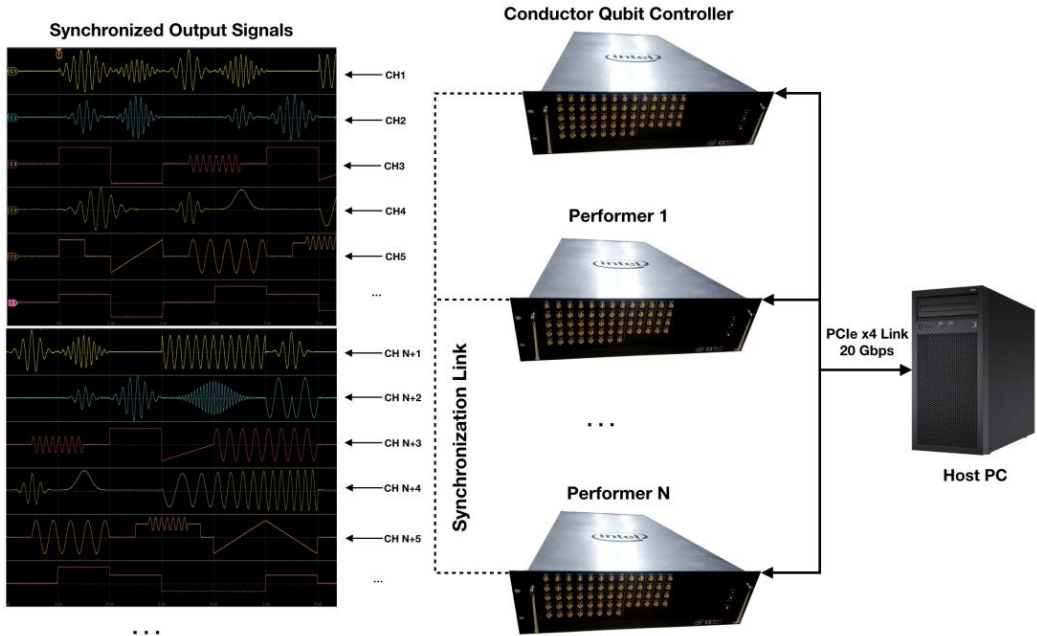

Figure 10. Multi-Controller System composed of a Conductor and several Performers. The various controllers are synchronized using a dedicated synchronization link and a communication protocol to ensure coherent qubit control.

For a given number of spin qubits ranging from few qubits to a hundred of qubits, Figure 11 shows the number of standard instruments needed to control the qubit setup in comparison with the number of Intel qubit controllers needed to control the same setup. The standard instruments are assumed to be arbitrary 8-Channels AWGs and standard Lock-In Amplifiers or Digitizers. The standard instruments are also assumed to be capable of synchronizing all the channels across different instruments similar to our qubit control system. The comparison is based on the number of required control channels for controlling single-electron spin qubits in gate-defined semiconductor quantum dots (QDs) [9][12].

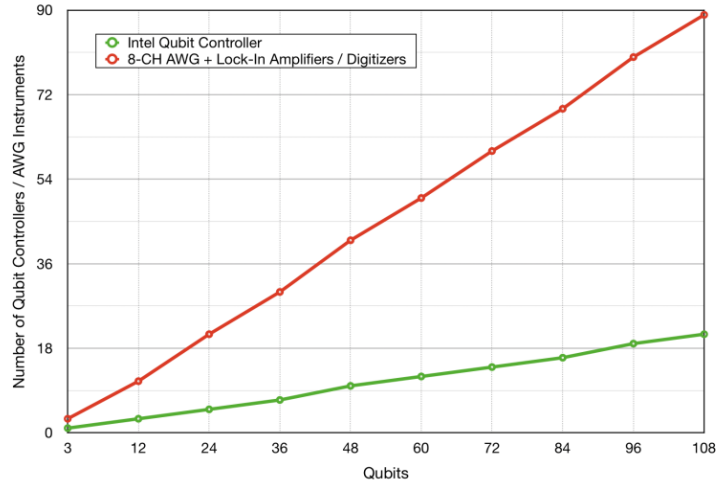

Figure 11. Comparison between the number of standard Instruments (AWGs and Lock-In Amplifiers) needed to control N qubits and the number of Intel Qubit Controllers required to control the same number of Qubits

# 6   Qubit Readout and Data Acquisition

Qubit readout is an essential component of a quantum computing system. Depending on the qubit technology, several readout techniques can be employed to measure one or more qubits simultaneously (e.g. using multiplexed readout). Our controller supports two different types of readout used by single-electron spin qubits:

- **Charge sensing** [18][19]: The incoming signal from the charge sensor is digitized and processed. The qubit state is discriminated by identifying whether a certain signal threshold has been crossed or not.
- **Reflectometry-based readout** [10][20]: This readout involves pulsing the resonance circuit attached to the sensor and processing the reflected signal. Impedance changes in the circuit correspond to the qubit state.

The result of a quantum computation is a discrete probability distribution on the possible qubit states. Hence, measurements are probabilistic in nature and requires sampling, averaging, and statistics. To facilitate this requirement the controller integrates a measurement averaging unit to automatically collect readout statistics, reducing the communication overhead with the software running on the host computer. Our qubit controller also allows digitization and storage of the raw readout signals so that specialized users can perform further signal analyses in software, if so desired.

# 7   Experimental Results: Coherent Spin Qubit Control

To verify the functionality of the qubit controller, we integrate it in a spin qubit measurement setup. To control spin qubits, a cryogenic dilution refrigerator with a base temperature of ≈ 10 mK is used. Voltage and microwave control lines allow operating a multi-gate double-channel device defined on a Si-SiGe heterostructure [14][12]. By shaping the electrostatic potential of the device using voltage gates, a QD is formed in each channel of the device. The DC voltages are fine-tuned to confine a single electron in one of the QDs. A strong magnetic field (B ≈ 0.4 T) is used to energy-split the spin states of the electron, which are used as the qubit |0> and |1> states. The other QD acts as a charge sensor, which can detect spin-selective charge tunneling of the single electron, and therefore can be used to perform readout of the qubit states [18]. With this readout scheme, a spin state visibility of ≈ 30% can be obtained, limited by sample performance.

The qubit control pulses are generated by the qubit controller. Crosstalk-corrected, fast voltage pulses move the qubit between its unload, load, manipulation, and readout points. Microwave tones, generated by upconverting shaped, frequency modulated control pulses, are used to manipulate the state of the qubit. The previously discussed quadrature modulation correction feature of our qubit controller is utilized to perform IQ phase, amplitude and offset correction. This ensures quality microwave modulation and eliminates unwanted spurs. Readout, including measurement outcome assignment of the individual measurement shots, is performed with the qubit controller. We want to note that we achieved comparable readout fidelity to commercially available lock-in amplifier while benefiting of faster readout signal processing, lower communication latency and the qubit readout outcome detection implemented in our qubit controller hardware.

To ensure coherent, high-fidelity operations, control pulses need to be carefully calibrated. Applying a microwave tone near the qubit resonance frequency induces Rabi oscillations between the |0> and |1>

states [Fig. 12(a-c)]. By varying frequency, duration and amplitude of the control pulse, optimal parameters for single qubit π and π/2 rotations can be determined.

With a Ramsey measurement, the coherence time T2* of the qubit is determined to be 1.2 µs [Fig. 12(d)], limited by nuclear spins of 29Si atoms in the vicinity of the QD. Extending the Ramsey to a Hahn- Echo measurement by applying a single refocusing pulse in the middle of the sequence makes the qubit insensitive to low-frequency noise and boosts the Echo coherence time T2Echo to 115 µs [Fig. 12(e)]. Further on, we demonstrate good qubit control by applying pairs of microwave control pulses In an AllXY measurement that bring the qubit state expectation value to 0, 0.5 or 1, respectively [Fig. 12(f)] [23]. These results demonstrate that our qubit controller is well capable of manipulating qubits in a highly coherent manner.

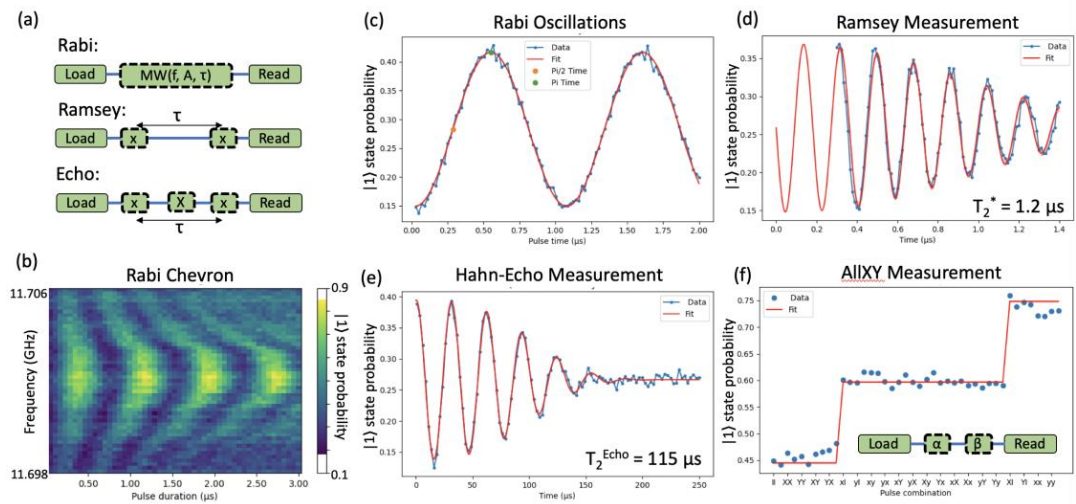

Figure 12: Coherent control of a spin qubit with the qubit controller. (a) Qubit control sequences: the electron is initialized in its |0> state, a sequence microwave control pulses is applied, and the qubit is read out. (b) Rabi chevron experiment, where the qubit expectation value is displayed as a function of microwave pulse frequency and duration. (c) Line-cut of (b) to extract optimal pulse time for π and π/2 rotations. (d, e) Ramsey and Hahn-Echo measurements demonstrate highly coherent qubit control. (f) Applying two qubit gates in an AllXY experiment to check pulse quality. Data points form a staircase pattern, showing well-tuned control pulses.

## 8   Conclusion and Future Work

In this paper we introduced a novel qubit control microarchitecture for efficient qubit control using a dedicated instruction set. We presented the new instruction set and showed how we can use direct digital synthesis to synthetize waveform on the fly and achieve high memory efficiency by avoiding the need for precomputing and storing waveforms. Besides the high programmability offered by our instruction set and the dynamic pulse synthesis capability, we described the integrated qubit readout implementation and the different readout modes supported by our controller such as charge sensing and reflectometry-based readout. We presented the implementation of our novel qubit control architecture on a Stratix FPGA platform and we described the built-in multi-controller mode that enables our qubit control system to scale to a larger number of qubits while maintaining synchronization across control units. This mode allows our qubit controller to achieve higher scalability in comparison with conventional instrument-based qubit control setups. In order to validate experimentally our qubit controller system, we demonstrated coherent spin qubit control. The system is well suited to perform qubit operations as demonstrated by measured high-quality single qubit gates and qubit readout. Two qubit gate has been also validated more recently to cover all elementary quantum operations.

Our qubit controller is being integrated in larger qubit systems, this will allow us to implement actual quantum algorithms and larger quantum workloads. To improve further the scalability of our qubit controller to control larger number of qubits, in the next steps, we will upgrade the multiplexing scheme of our control channels and increase their bandwidth to achieve higher control channel density. Moreover, implementing several calibration and tuning routines in our qubit controller hardware will help automating those routines and significantly reduce the calibration time especially on large qubit systems where tuning individual qubits manually in software can be very time consuming. That will also allow us to track the qubit control parameters while they shift over time and compensate for those shifts dynamically. Besides providing near-term qubit chip control at room temperature, our qubit controller provides an FPGA prototyping platform for developing a quantum microarchitecture that can be integrated in the digital design of cryogenic control systems such as in the Intel Horse Ridge cryo-controller chip [15] .